\documentclass[aps,twocolumn]{revtex4}

\usepackage{amsmath}
\usepackage{graphicx}

\usepackage{epsfig}

\def\ov#1{\overline{#1}}

\def\pd#1#2{\frac{\partial #1}{\partial #2}}

\def\cal#1{\mathcal{#1}}

\def\exd{{\sf d}}

\newcommand{\bc}{\begin{center}}
\newcommand{\ec}{\end{center}}
\newcommand{\bt}{\begin{tabbing}}
\newcommand{\et}{\end{tabbing}} 
\newcommand{\be}{\begin{eqnarray*}}
\newcommand{\ee}{\end{eqnarray*}}
\newcommand{\bs}{\begin{slide}}
\newcommand{\es}{\end{slide}}

\begin{document}

\title{Jacobi zeta function and action-angle coordinates for the pendulum}

\author{Alain J.~Brizard}
\affiliation{Department of Physics, Saint Michael's College, Colchester, VT 05439, USA} 

\begin{abstract}
The Jacobi elliptic functions and integrals play a defining role in analytically describing the motion of the planar pendulum. In the present paper, the Jacobi zeta function is given the physical interpretation as the generating function of the canonical transformation from the pendulum coordinates 
$\vartheta$ and $p \equiv \partial\vartheta/\partial t$ to the action-angle coordinates $(J,\zeta)$ for both the librating pendulum and the rotating pendulum.
\end{abstract}

\begin{flushright}
August 17, 2012
\end{flushright}

\maketitle

\section{\label{sec:intro}Introduction}

It is often said that, in order to understand the Jacobi elliptic functions, one only needs to study the pendulum problem \cite{Greenhill}. Jacobi elliptic functions are doubly-periodic functions that are useful in solving many problems in classical mechanics, general relativity and cosmology, as well as the traveling-soliton solutions of several nonlinear partial differential equations \cite{Lawden,Whittaker,Landau,Brizard,Brizard_EJP}. For an excellent introduction to the Jacobi elliptic functions see Ref.~\cite{Lawden} while a survey of the properties of elliptic functions and integrals is given in Refs.~\cite{NIST_EK,NIST_Jacobi}. The Jacobi functions are familiar to many physicists because they transform into the well-known singly-periodic trigonometric (or hyperbolic trigonometric) functions in the limit where the imaginary (or real) Jacobi period becomes infinite.

The Jacobi elliptic functions have recently been used to describe the guiding-center motion of trapped/passing particles in axisymmetric tokamak geometry \cite{Brizard_gc}. In particular, explicit formulas were provided for the canonical transformation to the bounce-center action-angle coordinates describing trapped/passing particle orbits. It was also noted that the problem of guiding-center particle orbits is strongly analogous to the pendulum problem (i.e., the trapped-particle guiding-center motion is analogous to the libration motion of a pendulum while the passing-particle guiding-center motion is analogous to the rotation motion of a pendulum). Hence, as a prelude to the search for the generating function for the bounce-center canonical transformation to action-angle coordinates, the purpose of the present paper is to find an explicit expression for the generating function of the canonical transformation to action-angle coordinates for the pendulum problem.

The pendulum problem is concerned with solutions of the (dimensionless) pendulum differential equation
\begin{equation}
\ddot{\vartheta} \;+\; \sin\vartheta \;=\; 0,
\label{eq:theta_ddot}
\end{equation}
which are subject to the initial conditions $(\vartheta_{0}, p_{0})$ chosen to satisfy the dimensionless energy equation
\begin{equation}
\frac{1}{2}\,p^{2} \;+\; (1 - \cos\vartheta) \;=\; \epsilon,
\label{eq:energy}
\end{equation}
where $\epsilon \equiv 2\,\kappa$ denotes the dimensionless pendulum energy and $p \equiv \partial\vartheta/\partial t$ denotes the pendulum (angular) momentum. The libration motion of the pendulum is defined for $\epsilon < 2$ while the rotation motion of the pendulum is defined for $\epsilon > 
2$. In the present paper, we show that the Jacobi zeta function \cite{Lawden} plays a fundamental role in generating the canonical transformation from the pendulum coordinates $(p,\vartheta)$ to the action-angle coordinates $(J,\zeta)$ for both the librating and rotating motions. 

We begin with a brief review of the solutions of librating and rotating motion of the pendulum problem expressed in terms of the Jacobi elliptic functions \cite{Lawden,Whittaker,Landau,Brizard}. The solution for the problem of the libration motion of a pendulum $(\epsilon < 2)$ is expressed in terms of the Jacobi elliptic functions ${\rm sn}(t|\kappa)$ and ${\rm cn}(t|\kappa)$ \cite{footnote} as
\begin{eqnarray}
\vartheta_{\ell}(t,\kappa) & = & 2\,\arcsin[\sqrt{\kappa}\;{\rm sn}(t|\kappa)], 
\label{eq:theta_lib} \\
p_{\ell}(t,\kappa) & = & 2\,\sqrt{\kappa}\;{\rm cn}(t|\kappa), 
\label{eq:p_lib}
\end{eqnarray}
where $t$ denotes the dimensionless time and the Jacobi parameter $\kappa$ is used to define the dimensionless energy of the librating pendulum ($\kappa < 1$). In the low-energy (small-amplitude) limit $\kappa \ll 1$, we recover the simple-pendulum solution $\vartheta_{\ell}(t,\kappa) \simeq 2\,\sqrt{\kappa}\,\sin t$ from the librating solution \eqref{eq:theta_lib}, where ${\rm sn}(t|\kappa) \simeq \sin t$ when $\kappa \ll 1$. 

For the rotation motion of the pendulum $(\epsilon > 2)$, the Jacobi elliptic functions can be evaluated according to the relations $(\kappa > 1)$
\begin{equation}
\left. \begin{array}{rcl}
\sqrt{\kappa}\;{\rm sn}(t|\kappa) & \equiv & {\rm sn}\left(\sqrt{\kappa}\,t|\kappa^{-1}\right)\\
 &  & \\
{\rm cn}(t|\kappa) & \equiv & {\rm dn}\left(\sqrt{\kappa}\,t|\kappa^{-1}\right) \\
 &  & \\
{\rm dn}(t|\kappa) & \equiv & {\rm cn}\left(\sqrt{\kappa}\,t|\kappa^{-1}\right)
\end{array} \right\},
\label{eq:cn_rot}
\end{equation}
and Eqs.~\eqref{eq:theta_lib}-\eqref{eq:p_lib} are replaced with
\begin{eqnarray}
\vartheta_{r}(t,\kappa) & = & 2\,\arcsin\left[{\rm sn}\left(\sqrt{\kappa}\,t|\kappa^{-1}\right)\right], \label{eq:theta_rot} \\
p_{r}(t,\kappa) & = & 2\,\sqrt{\kappa}\;{\rm dn}\left(\sqrt{\kappa}\,t|\kappa^{-1}\right). \label{eq:p_rot}
\end{eqnarray}
Note that the librating pendulum oscillates between the turning points $\pm\,\vartheta_{\ell0}(\kappa)$, where
\begin{equation}
0 \;\leq\; \vartheta_{\ell0}(\kappa) \;\equiv\; 2\,\arcsin(\sqrt{\kappa}) \;<\; \pi, 
\label{eq:theta_bounce}
\end{equation}
while the range of motion for the rotating pendulum is $-\,\pi \leq \vartheta_{r} \leq \pi$ (where $-\,\pi$ and $\pi$ are now considered to be identical points).

\begin{figure}
\epsfysize=2in
\epsfbox{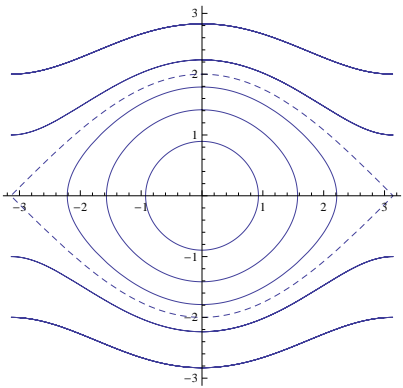}
\caption{Phase portrait $(\vartheta(t), p(t))$ of the pendulum motion showing the librating solution \eqref{eq:theta_lib}-\eqref{eq:p_lib} and the rotating solution \eqref{eq:theta_rot}-\eqref{eq:p_rot}, for $-\,\pi \leq \vartheta \leq \pi$. The separatrix solution \eqref{eq:thetap_sep}, which separates the librating solution (inside) from the rotating solution (outside), is shown as a dashed line.}
\label{fig:pendulum} 
\end{figure}

Using the solutions \eqref{eq:theta_lib}-\eqref{eq:p_lib} and \eqref{eq:theta_rot}-\eqref{eq:p_rot}, and the trigonometric identity $1 - \cos\vartheta = 
2\,\sin^{2}(\vartheta/2)$, we note that the energy equation \eqref{eq:energy} follows from the Jacobi identities
\begin{equation}
\left. \begin{array}{r}
{\rm cn}^{2}(t|\kappa) + {\rm sn}^{2}(t|\kappa) \\ 
 \\
{\rm dn}^{2}\left(\sqrt{\kappa}\,t|\kappa^{-1}\right) + \kappa^{-1}\;{\rm sn}^{2}\left(\sqrt{\kappa}\,t|\kappa^{-1}\right) 
\end{array} \right\} \;=\; 1
\label{lib_rot_id}
\end{equation}
for the librating and rotating solutions, respectively.

Lastly, the separatrix solution (which separates the librating solution from the rotating solution) is obtained by substituting the value $\kappa = 1$ in Eqs.~\eqref{eq:theta_lib}-\eqref{eq:p_lib} [or Eqs.~\eqref{eq:theta_rot}-\eqref{eq:p_rot}]:
\begin{equation}
\left. \begin{array}{rcl}
\vartheta_{s}(t) & = & 2\;\arcsin(\tanh\,t) \\
p_{s}(t) & = & 2\,{\rm sech}\,t
\end{array} \right\}, 
\label{eq:thetap_sep}
\end{equation}
where we used ${\rm sn}(t|1) = \tanh\,t$ and ${\rm cn}(t|1) = {\rm dn}(t|1) = {\rm sech}\,t$. Note that the turning points $\pm\,\pi$ of the separatrix solution \eqref{eq:thetap_sep} are now reached only after an infinite period of time.

The remainder of this paper is organized as follows. In Sec.~\ref{sec:action_angle}, we introduce the action-angle coordinates $(J,\zeta)$ for each type of pendulum motion, and the librating and rotating solutions are reformulated in terms of Jacobi elliptic functions of $(J,\zeta)$. In Sec.~\ref{sec:canonical}, we show that the transformation $(p,\vartheta) \rightarrow (J,\zeta)$ is a canonical transformation and we find the generating function $S$ for this transformation for each type of motion in Sec.~\ref{sec:generate}. Here, we show that the Jacobi zeta function appears naturally as the generating function for the canonical transformation $(p,\vartheta) \rightarrow (J,\zeta)$ for both types of motion. 

\section{\label{sec:action_angle}Action-angle Coordinates}

In this Section, we calculate explicit expressions for the action-angle coordinates associated with the pendulum problem discussed in 
Sec.~\ref{sec:intro}. We begin with the action coordinate
\begin{equation}
J(\kappa) \;\equiv\; \frac{1}{2\pi}\;\oint\,p(\vartheta,\kappa)\,d\vartheta
\label{eq:J_def}
\end{equation}
for the librating and rotating motions of the pendulum, where the magnitude of the momentum
\begin{equation}
|p|(\vartheta, \kappa) \;=\; 2\;\sqrt{\kappa \;-\; \sin^{2}(\vartheta/2)}
\label{eq:p_theta}
\end{equation}
vanishes at $\vartheta = \pm\,\vartheta_{\ell0}(\kappa)$ for the librating case $(\kappa < 1)$ [see Eq.~\eqref{eq:theta_bounce}] while the motion of a co-rotating pendulum $(p > 0)$ is separated from that of a counter-rotating pendulum $(p < 0)$.

Once the action coordinates $J_{\ell}(\kappa)$ and $J_{r}(\kappa)$ have been calculated as functions of energy $\epsilon = 2\,\kappa$, we can calculate the pendulum frequencies $\omega \equiv (\partial J/\partial\epsilon)^{-1}$. Lastly, by defining the angle coordinates $\zeta_{\ell} \equiv \omega_{\ell}\,t$ and $\zeta_{r} \equiv \omega_{r}\,t$, the librating solutions \eqref{eq:theta_lib}-\eqref{eq:p_lib} and the rotating solutions \eqref{eq:theta_rot}-\eqref{eq:p_rot} can be expressed in terms of their respective action-angle coordinates.

\subsection{Libration Case}

The action coordinate for the librating pendulum is calculated from Eq.~\eqref{eq:J_def} as
\begin{eqnarray}
J_{\ell}(\kappa) & = & \frac{4\kappa}{\pi}\,\int_{-\pi/2}^{\pi/2}\;\frac{\cos^{2}\phi\;d\phi}{
\sqrt{1 - \kappa\,\sin^{2}\phi}} \nonumber \\
 & = & \frac{8}{\pi} \left[ {\sf E}(\kappa) \;-\frac{}{} (1 - \kappa)\,{\sf K}(\kappa) \right],
\label{eq:J_lib}
\end{eqnarray}
where we used the substitution $\sin(\vartheta/2) = \sqrt{\kappa}\,\sin\phi$ in Eq.~\eqref{eq:p_theta}, while ${\sf K}(\kappa)$ and ${\sf E}(\kappa)$ denote the complete elliptic integrals of the first and second kinds \cite{NIST_EK}. In the low-energy limit $(\kappa \ll 1)$, Eq.~\eqref{eq:J_lib} yields $J_{\ell}(\kappa) \simeq 2\,\kappa$. 

The librating-pendulum frequency is defined from Eq.~\eqref{eq:J_lib} as
\begin{equation}
\omega_{\ell}(\kappa) \;\equiv\; \left(\frac{1}{2}\,\pd{J_{\ell}(\kappa)}{\kappa}\right)^{-1} \;=\; \frac{\pi}{2\,{\sf K}(\kappa)},
\label{eq:omega_lib}
\end{equation}
where we used the expression
\[ \frac{d}{d\kappa} \left[ {\sf E}(\kappa) \;-\frac{}{} (1 - \kappa)\,{\sf K}(\kappa) \right] \;=\; \frac{1}{2}\;
{\sf K}(\kappa). \]
In the low-energy limit $(\kappa \ll 1)$, Eq.~\eqref{eq:omega_lib} yields $\omega_{\ell} \simeq 1$ (i.e., $T_{\ell} \simeq 
2\pi$). 

The librating solutions \eqref{eq:theta_lib}-\eqref{eq:p_lib} can now be re-expressed in terms of the action-angle 
$(J_{\ell},\zeta_{\ell})$ coordinates as
\begin{eqnarray}
\vartheta_{\ell}(J_{\ell},\zeta_{\ell}) & = & 2\,\arcsin\left[\sqrt{\kappa}\;{\rm sn}\left(\frac{2{\sf K}}{\pi}\,\zeta_{\ell} \,\left.\frac{}{}\right|\;\kappa\right)\right],
\label{eq:theta_lib_zetaJ} \\
p_{\ell}(J_{\ell},\zeta_{\ell}) & = & 2\sqrt{\kappa}\;{\rm cn}\left(\frac{2{\sf K}}{\pi}\,\zeta_{\ell} \,\left.\frac{}{}\right|\;\kappa\right),
\label{eq:p_lib_zetaJ}
\end{eqnarray}
where the libration angle $\zeta_{\ell}$ is defined as
\begin{equation}
\zeta_{\ell} \;\equiv\; \omega_{\ell}\;t \;=\; (\pi/2)\,t/{\sf K}(\kappa),
\label{eq:zeta_lib}
\end{equation}
and $\kappa(J_{\ell})$ is obtained from Eq.~\eqref{eq:J_lib}. In the low-energy limit $(\kappa \ll 1)$, 
Eqs.~\eqref{eq:theta_lib_zetaJ}-\eqref{eq:p_lib_zetaJ} become
\begin{equation}
\left. \begin{array}{rcl}
\vartheta_{\ell}(J_{\ell},\zeta_{\ell}) & \simeq & 2\,\sqrt{\kappa}\;\sin\zeta_{\ell} \\
 &  & \\
p_{\ell}(J_{\ell},\zeta_{\ell}) &  \simeq & 2\,\sqrt{\kappa}\;\cos\zeta_{\ell}
\end{array} \right\},
\label{eq:thetap_low}
\end{equation}
where $J_{\ell} \simeq 2\,\kappa$ and $\zeta_{\ell} \simeq t$. 

\subsection{Rotation Case}

Next, the action coordinate for the co-rotating pendulum $(p > 0)$ is calculated from Eq.~\eqref{eq:J_def} as
\begin{eqnarray}
J_{r}(\kappa) & = & \frac{2\sqrt{\kappa}}{\pi}\,\int_{-\pi/2}^{\pi/2}\;\sqrt{1 - \kappa^{-1}\,
\sin^{2}\phi}\;d\phi \nonumber \\
 & = &  \frac{4\sqrt{\kappa}}{\pi}\,{\sf E}\left(\kappa^{-1}\right),
\label{eq:J_rot}
\end{eqnarray}
where we used the substitution $\vartheta = 2\,\phi$ in Eq.~\eqref{eq:p_theta}; the action of a counter-rotating pendulum $(p < 0)$ has the value 
$-\,J_{r}(\kappa)$ at the same energy $\epsilon = 2\,\kappa$. Note that the action coordinate is discontinuous at the separatrix $(\kappa = 1)$: 
$J_{\ell}(1) = 8/\pi$ and $J_{r}(1) = 4/\pi$, where we used the limits ${\sf E}(1) = 1$ and $\lim_{\kappa \rightarrow 1}(1 - \kappa)\,{\sf K}(\kappa) =
 0$ in Eqs.~\eqref{eq:J_lib} and \eqref{eq:J_rot}. The extra factor of 2 comes from the fact that the full librating cycle requires a passage from $-\,\vartheta_{\ell0}(\kappa)$ to $\vartheta_{\ell0}(\kappa)$ and back to $-\,\vartheta_{\ell0}(\kappa)$, while the full rotating cycle only requires a one-way passage from $-\,\pi$ to $\pi$ (which is identical to $-\,\pi$).

The rotating-pendulum frequency, on the other hand, is defined from Eq.~\eqref{eq:J_rot} as
\begin{equation}
\omega_{r}(\kappa) \;\equiv\; \left(\frac{1}{2}\,\pd{J_{r}(\kappa)}{\kappa}\right)^{-1} \;=\; \frac{\pi\,\sqrt{\kappa}}{{\sf K}(\kappa^{-1})},
\label{eq:omega_rot}
\end{equation}
where we used the expression
\[ \frac{d}{d\kappa} \left[ \sqrt{\kappa}\frac{}{}{\sf E}\left( \kappa^{-1}\right)\right] \;=\; \frac{{\sf K}(\kappa^{-1})}{2\,\sqrt{\kappa}}. \]
By comparing Eqs.~\eqref{eq:J_lib} and \eqref{eq:J_rot} with Eqs.~\eqref{eq:omega_lib} and \eqref{eq:omega_rot}, respectively, we obtain the relations among complete elliptic integrals \cite{NIST_EK}:
\begin{eqnarray}
{\cal Re}[{\sf K}(\kappa)] & = & {\sf K}(\kappa^{-1})/\sqrt{\kappa}, \label{eq:K_inverse} \\
{\cal Re}[{\sf E}(\kappa)] & = & \sqrt{\kappa}\left[{\sf E}(\kappa^{-1}) \;-\frac{}{} (1 - \kappa^{-1})\;{\sf K}(\kappa^{-1})\right], \label{eq:E_inverse}
\end{eqnarray}
where $\kappa > 1$ and ${\cal Re}[\cdots]$ denotes the real part. Hence, the pendulum problem naturally gives the extension of the complete elliptic integrals ${\sf E}$ and ${\sf K}$ for $\kappa > 1$. As discussed with the separatrix solution \eqref{eq:thetap_sep}, the frequencies \eqref{eq:omega_lib}-\eqref{eq:omega_rot} go to zero (i.e., the periods become infinite) in the limit $\kappa \rightarrow 1$ (since ${\sf K}$ goes to infinity in that limit).

The rotating solutions \eqref{eq:theta_rot}-\eqref{eq:p_rot} can also be re-expressed in terms of the action-angle $(J_{r},\zeta_{r})$ coordinates as
\begin{eqnarray}
\vartheta_{r}(J_{r},\zeta_{r}) & = & 2\,\arcsin\left[{\rm sn}\left({\sf K}(\kappa^{-1})\,\frac{\zeta_{r}}{\pi} \,\left.\frac{}{}\right|\,\kappa^{-1}
\right)\right],
\label{eq:theta_rot_zetaJ} \\
p_{r}(J_{r},\zeta_{r}) & = & 2\sqrt{\kappa}\;{\rm dn}\left({\sf K}(\kappa^{-1})\,\frac{\zeta_{r}}{\pi} \,\left.\frac{}{}\right|\,\kappa^{-1}\right),
\label{eq:p_rot_zetaJ}
\end{eqnarray}
where the rotation angle $\zeta_{r}$ is defined as
\begin{equation}
\zeta_{r} \;\equiv\; \omega_{r}\;t \;=\; \pi\,\sqrt{\kappa}\,t/{\sf K}(\kappa^{-1}),
\label{eq:zeta_rot}
\end{equation}
and $\kappa(J_{r})$ is obtained from Eq.~\eqref{eq:J_rot}. 

\section{\label{sec:canonical}Canonical Transformation}

We now show that the phase-space transformation 
\begin{equation}
(\vartheta,p) \;\rightarrow\; (\zeta,J) 
\label{eq:canonical_trans}
\end{equation}
is canonical for both the librating solution \eqref{eq:theta_lib_zetaJ}-\eqref{eq:p_lib_zetaJ} and the rotating solution 
\eqref{eq:theta_rot_zetaJ}-\eqref{eq:p_rot_zetaJ} by proving the canonical relation
\begin{equation}
\pd{\vartheta}{\zeta}\,\pd{p}{J} \;-\; \pd{\vartheta}{J}\,\pd{p}{\zeta} \;=\; 1
\label{eq:canonical}
\end{equation}
for each solution. 

\subsection{Libration Case}

We first consider the case of the librating pendulum $(\kappa < 1)$. For this case (where $t = \zeta/\omega_{\ell}$), the partial derivatives in Eq.~\eqref{eq:canonical} are
\begin{eqnarray}
\left.\pd{}{\zeta}\right|_{J} & \equiv & \frac{1}{\omega_{\ell}}\;\pd{}{t}, \label{eq:partial_zeta_t} \\
\left.\pd{}{J}\right|_{\zeta} & \equiv & \frac{\omega_{\ell}}{2} \left( \pd{}{\kappa} \;+\; 
\frac{1}{\Omega_{\ell}}\;\pd{}{t} \right), \label{eq:partial_J_kappa}
\end{eqnarray}
where $\partial/\partial t$ and $\partial/\partial\kappa$ are understood to be at constant $\kappa$ and $t$, respectively, and we have introduced the definition
\begin{equation}
\frac{1}{\Omega_{\ell}} \;\equiv\; \left.\pd{t}{\kappa}\right|_{\zeta} \;=\; \frac{t}{2\kappa\,(1 - \kappa)}\;\left[
\frac{\sf E}{\sf K} \;-\; (1 - \kappa) \right].
\label{eq:Omega_lib_def}
\end{equation}
We thus use Eqs.~\eqref{eq:theta_lib_zetaJ}-\eqref{eq:p_lib_zetaJ} and \eqref{eq:partial_zeta_t} to obtain from Eqs.~\eqref{eq:theta_lib}-\eqref{eq:p_lib}:
\begin{eqnarray}
\pd{\vartheta}{\zeta} \;=\; \frac{1}{\omega_{\ell}}\,\pd{\vartheta}{t} & = & \frac{2\,\sqrt{\kappa}}{\omega_{\ell}}\;{\rm cn}, \label{eq:vartheta_t_lib} \\
\pd{p}{\zeta} \;=\; \frac{1}{\omega_{\ell}}\,\pd{p}{t} & = & -\,\frac{2\sqrt{\kappa}}{\omega_{\ell}}\;{\rm sn}\,{\rm dn},
\label{eq:p_t_lib}
\end{eqnarray}
and then, using Eq.~\eqref{eq:partial_J_kappa}, we obtain
\begin{eqnarray}
\pd{\vartheta}{J} & = & \frac{\omega_{\ell}}{2\,\sqrt{\kappa}}\,\left[ {\rm sd} \;+\; \frac{2\kappa}{{\rm dn}}\left(
\pd{{\rm sn}}{\kappa} \;+\; \frac{1}{\Omega_{\ell}}\;\pd{{\rm sn}}{t} \right) \right], \label{eq:vartheta_kappa_lib} \\
\pd{p}{J} & = & \frac{\omega_{\ell}}{2\,\sqrt{\kappa}}\,\left[ {\rm cn} \;+\; 2\kappa\;\left(
\pd{{\rm cn}}{\kappa} \;+\; \frac{1}{\Omega_{\ell}}\;\pd{{\rm cn}}{t} \right) \right],
\label{eq:p_kappa_lib} 
\end{eqnarray}
where ${\rm sd} \equiv {\rm sn}/{\rm dn}$ (following standard notation for Jacobi elliptic functions). By inserting these expressions into 
Eq.~\eqref{eq:canonical}, we obtain
\begin{equation}
\pd{\vartheta}{\zeta}\,\pd{p}{J} \;-\; \pd{\vartheta}{J}\,\pd{p}{\zeta} \;=\; \pd{}{\kappa} \left[ \kappa\frac{}{} \left(
{\rm cn}^{2} + {\rm sn}^{2} \right) \right] \;=\; 1,
\label{eq:canonical_proof_lib}
\end{equation}
which proves the canonical relation \eqref{eq:canonical} for the librating case when the identity \eqref{lib_rot_id} is used.

\begin{figure}
\epsfysize=2in
\epsfbox{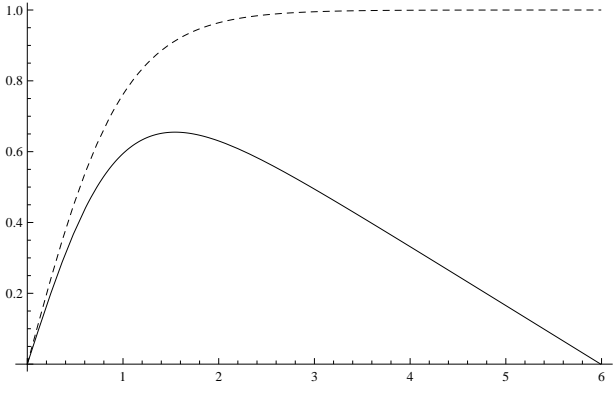}
\caption{Plot of ${\rm Z}(t|\kappa)$ versus $t$ in the range $0 \leq t \leq {\sf K}(\kappa)$ for $\kappa = 0.9999$ (solid) and ${\rm Z}(t|1) = \tanh t$ (dashed).}
\label{fig:zeta} 
\end{figure}

In writing Eqs.~\eqref{eq:vartheta_kappa_lib}-\eqref{eq:p_kappa_lib} explicitly, the partial derivatives
\begin{eqnarray}
\pd{{\rm sn}}{\kappa} \;+\; \frac{1}{\Omega_{\ell}}\;\pd{\rm sn}{t} & = & \frac{\kappa\,{\rm sn}\,{\rm cn}^{2} \;-\; 
{\rm Z}\;{\rm cn}\,{\rm dn}}{2\,\kappa\,(1 - \kappa)}, 
\label{eq:partial_sn} \\
\pd{{\rm cn}}{\kappa} \;+\; \frac{1}{\Omega_{\ell}}\;\pd{\rm cn}{t} & = & \frac{{\rm Z}\;{\rm sn}\,{\rm dn} \;-\; \kappa\,
{\rm cn}\,{\rm sn}^{2}}{2\,\kappa\,(1 - \kappa)}, 
\label{eq:partial_cn}
\end{eqnarray}
are expressed in terms of the Jacobi zeta function \cite{Lawden}
\begin{equation}
{\rm Z}(t|\kappa) \;\equiv\; \int_{0}^{t} \left( {\rm dn}^{2}(s|\kappa) \;-\; \frac{{\sf E}}{{\sf K}} \right)\;ds.
\label{eq:Zeta_def}
\end{equation}
The Jacobi zeta function \eqref{eq:Zeta_def} has odd parity: ${\rm Z}(-t|\kappa) = -\,{\rm Z}(t|\kappa)$, it has a period of 
$2\,{\sf K}(\kappa)$: ${\rm Z}(t + 2\,{\sf K}|\kappa) \;=\; {\rm Z}(t|\kappa)$, and it vanishes at $t = n{\sf K}$, with $n = 0, \pm\,1, \pm\,2, \cdots$: ${\rm Z}(n{\sf K}|\kappa) = 0$. The Jacobi zeta function \eqref{eq:Zeta_def} is also evaluated for the separatrix case $(\kappa = 1)$ as
\begin{equation}
{\rm Z}(t|1) \;=\; \int_{0}^{t} {\rm sech}^{2}s\;ds \;=\; \tanh\,t.
\label{eq:Z_1}
\end{equation}
Figure \ref{fig:zeta} shows the plot of ${\rm Z}(t|\kappa)$ for $\kappa = 0.9999$ (solid) and the plot of ${\rm Z}(t|1) = \tanh t$ (dashed). In the low-energy limit $(\kappa \ll 1)$,  we use ${\rm dn}^{2}(s|\kappa) - {\sf E}/{\sf K} \simeq \frac{1}{2}\,\kappa\;\cos\,2s$ in Eq.~\eqref{eq:Zeta_def}, which yields $Z(t|\kappa) \simeq \frac{1}{4}\,\kappa\;\sin\,2t$.

Lastly, using Eqs.~\eqref{eq:partial_sn}-\eqref{eq:partial_cn}, Eqs.~\eqref{eq:vartheta_kappa_lib}-\eqref{eq:p_kappa_lib} become
\begin{eqnarray}
\pd{\vartheta}{J} & = & \frac{\omega_{\ell}}{2\,\sqrt{\kappa}\,(1 - \kappa)}\;\left( {\rm sn}\,{\rm dn} \;-\frac{}{}
{\rm Z}\,{\rm cn} \right), \label{eq:vartheta_J_lib} \\
\pd{p}{J} & = & \frac{\omega_{\ell}}{2\,\sqrt{\kappa}\,(1 - \kappa)} \left[ {\rm cn}\,\left( {\rm dn}^{2} - \kappa\right) 
\;+\frac{}{} {\rm Z}\,{\rm sn}\,{\rm dn} \right], \label{eq:p_J_lib}
\end{eqnarray}
and we verify again that the canonical relation \eqref{eq:canonical} is satisfied.

\subsection{Rotation Case}

Next, we consider the case of the rotating pendulum $(\kappa > 1)$. For this case (where $u \equiv \sqrt{\kappa}\,t = 
\sqrt{\kappa}\;\zeta/\omega_{r}$), the partial derivatives in Eq.~\eqref{eq:canonical} are
\begin{eqnarray}
\left.\pd{}{\zeta}\right|_{J} & \equiv & \frac{\sqrt{\kappa}}{\omega_{r}}\;\pd{}{u}, \label{eq:partial_zeta_u} \\
\left.\pd{}{J}\right|_{\zeta} & \equiv & \frac{\omega_{r}}{2} \left( \pd{}{\kappa} \;+\; \frac{1}{\Omega_{r}}\;\pd{}{u} 
\right), \label{eq:partial_J_kappa_u}
\end{eqnarray}
where $\partial/\partial u$ and $\partial/\partial\kappa$ are understood to be at constant $\kappa$ and $u$, respectively, and we have introduced the definition
\begin{equation}
\frac{1}{\Omega_{r}} \;\equiv\; \left.\pd{u}{\kappa}\right|_{\zeta} \;=\; -\;\frac{u}{2\,(\kappa-1)}\;\left[
\frac{\ov{\sf E}}{\ov{\sf K}} \;-\; \left(1 - \kappa^{-1}\right) \right],
\label{eq:Omega_rot_def}
\end{equation}
where $\ov{\sf E} \equiv {\sf E}(\kappa^{-1})$ and $\ov{\sf K} \equiv {\sf K}(\kappa^{-1})$. We thus use Eqs.~\eqref{eq:theta_rot_zetaJ}-\eqref{eq:p_rot_zetaJ} and \eqref{eq:partial_zeta_u} to obtain from Eqs.~\eqref{eq:theta_rot}-\eqref{eq:p_rot}:
\begin{eqnarray}
\pd{\vartheta}{\zeta} & = & \frac{2\,\sqrt{\kappa}}{\omega_{r}}\;\ov{\rm dn}, 
\label{eq:vartheta_t_rot} \\
\pd{p}{\zeta} & = & -\,\frac{2}{\omega_{r}}\;\ov{\rm sn}\,\ov{\rm cn}, \label{eq:p_t_rot}
\end{eqnarray}
where we use the overbar notation $\ov{\rm pq} \equiv {\rm pq}(u|\kappa^{-1})$, and then, using Eqs.~\eqref{eq:theta_rot_zetaJ}-\eqref{eq:p_rot_zetaJ}, we obtain
\begin{eqnarray}
\pd{\vartheta}{J} & = & \frac{\omega_{r}}{\ov{\rm cn}} \left( \pd{\ov{\rm sn}}{\kappa} \;+\; \frac{1}{\Omega_{r}}\,
\pd{\ov{\rm sn}}{u} \right), \label{eq:vartheta_kappa_rot} \\
\pd{p}{J} & = & \frac{\omega_{r}}{2\,\sqrt{\kappa}} \left[ \ov{\rm dn} \;+\; 2\kappa\;\left( \pd{\ov{\rm dn}}{\kappa} \;+\; 
\frac{1}{\Omega_{r}}\,\pd{\ov{\rm dn}}{u} \right) \right]. 
\label{eq:p_kappa_rot}
\end{eqnarray}
By inserting these expressions into Eq.~\eqref{eq:canonical}, we obtain
\begin{equation}
\pd{\vartheta}{\zeta}\,\pd{p}{J} - \pd{\vartheta}{J}\,\pd{p}{\zeta} = \pd{}{\kappa} \left[ \kappa\frac{}{} \left(
\ov{\rm dn}^{2} + \kappa^{-1}\ov{\rm sn}^{2} \right) \right] = 1,
\label{eq:canonical_proof_rot}
\end{equation}
which proves the canonical relation \eqref{eq:canonical} for the rotating case when the identity \eqref{lib_rot_id} is used.

In writing Eqs.~\eqref{eq:vartheta_kappa_rot}-\eqref{eq:p_kappa_rot}, the partial derivatives
\begin{eqnarray}
\pd{\ov{\rm sn}}{\kappa} \;+\; \frac{1}{\Omega_{r}}\;\pd{\ov{\rm sn}}{u} & = & \frac{\ov{\rm Z}\;\ov{\rm cn}\,\ov{\rm dn}
\;-\; \kappa^{-1}\ov{\rm sn}\,\ov{\rm cn}^{2}}{2\,(\kappa - 1)}, 
\label{eq:partial_sn_u} \\
\pd{\ov{\rm dn}}{\kappa} \;+\; \frac{1}{\Omega_{r}}\;\pd{\ov{\rm dn}}{u} & = & \frac{\ov{\rm dn}\,\ov{\rm sn}^{2} \;-\;
\ov{\rm Z}\;\ov{\rm sn}\,\ov{\rm cn}}{2\,\kappa\,(\kappa - 1)}, 
\label{eq:partial_dn_u}
\end{eqnarray}
are expressed in terms of the Jacobi zeta function \cite{Lawden}
\begin{equation}
\ov{\rm Z} \;\equiv\; {\rm Z}(u|\kappa^{-1}) \;=\; \int_{0}^{u} \left( {\rm dn}^{2}(s|\kappa^{-1}) \;-\; 
\frac{\ov{\sf E}}{\ov{\sf K}} \right)\;ds.
\label{eq:Zeta_def_u}
\end{equation}
Using 
Eqs.~\eqref{eq:partial_sn_u}-\eqref{eq:partial_dn_u}, Eqs.~\eqref{eq:vartheta_kappa_rot}-\eqref{eq:p_kappa_rot} become
\begin{eqnarray}
\pd{\vartheta}{J} & = & \frac{\omega_{r}}{2\,(\kappa - 1)}\;\left( \ov{\rm Z}\,\ov{\rm dn} \;-\frac{}{} \kappa^{-1}
\ov{\rm sn}\,\ov{\rm cn}  \right), \label{eq:vartheta_J_rot} \\
\pd{p}{J} & = & \frac{\omega_{r}}{2\,\sqrt{\kappa}\,(\kappa - 1)} \left[ \ov{\rm dn}\,\left( \kappa - \ov{\rm cn}^{2}\right) 
\;-\frac{}{} \ov{\rm Z}\,\ov{\rm sn}\,\ov{\rm cn} \right], \label{eq:p_J_rot}
\end{eqnarray}
which again satisfies the canonical relation \eqref{eq:canonical}. 

\section{\label{sec:generate}Generating function}

Now that we have established the canonical nature of the phase-space transformation \eqref{eq:canonical_trans}, we seek its generating function. First, we note that the canonical relation \eqref{eq:canonical} can also be expressed as the differential two-form identity
\begin{equation}
\exd p\;\wedge\;\exd\vartheta \;\equiv\; \exd J\;\wedge\;\exd\zeta,
\label{eq:canonical_twoform}
\end{equation} 
where $\exd$ denotes an exterior derivative (with $\exd^{2} \equiv 0$). The differential canonical relation 
\eqref{eq:canonical_twoform} therefore allows us to write the one-form $p\,\exd\vartheta$ in action-angle space as
\begin{equation}
p\;\exd\vartheta \;\equiv\; J\;\exd\zeta \;+\; \exd S,
\label{eq:S_oneform}
\end{equation}
where the function $S(\zeta,J)$ generates the canonical transformation \eqref{eq:canonical_trans}, and the identity $\exd^{2}S \equiv 0$ guarantees the canonical relation \eqref{eq:canonical_twoform}. Specifically, the generating function $S(\zeta,J)$ must satisfy the partial derivatives
\begin{eqnarray}
\pd{S}{\zeta} & \equiv & p\;\pd{\vartheta}{\zeta} \;-\; J, \label{eq:S_zeta_def} \\
\pd{S}{J} & \equiv & p\;\pd{\vartheta}{J}, \label{eq:S_J_def}
\end{eqnarray}
for the libration and rotation cases, which we now investigate separately. 

\subsection{Libration Case}

First, we consider the librating-pendulum case, represented by Eqs.~\eqref{eq:theta_lib_zetaJ}-\eqref{eq:p_lib_zetaJ},
\eqref{eq:vartheta_t_lib}-\eqref{eq:p_t_lib}, and \eqref{eq:vartheta_J_lib}-\eqref{eq:p_J_lib}. Equation \eqref{eq:S_zeta_def} for the libration case can be explicitly written as
\begin{equation}
\pd{S_{\ell}}{\zeta} \;=\; \frac{1}{\omega_{\ell}}\;\pd{S_{\ell}}{t} \;=\; \frac{4\kappa}{\omega_{\ell}}\;{\rm cn}^{2} \;-\; J_{\ell},
\label{eq:S_zeta_lib}
\end{equation}
where Eqs.~\eqref{eq:p_lib_zetaJ} and \eqref{eq:vartheta_t_lib} were used on the right side. Equation \eqref{eq:S_zeta_lib} can be integrated with respect to $t$ to give
\begin{equation}
S_{\ell}(t,\kappa) \;=\; 4\kappa\;\int_{0}^{t}\;{\rm cn}^{2}(s|\kappa)\,ds \;-\; J_{\ell}\,\zeta_{\ell},
\label{eq:S_first_lib}
\end{equation}
where we assumed that $S_{\ell} = 0$ at $t = 0$.

Next, we use the identity $\kappa\,{\rm cn}^{2} = {\rm dn}^{2} - (1 - \kappa)$, and the definition \eqref{eq:Zeta_def} for the Jacobi zeta function, to obtain
\begin{eqnarray}
\int_{0}^{t}\,\kappa\,{\rm cn}^{2}(s|\kappa)\,ds & = & Z(t|\kappa) \;+\; \left[ {\sf E} - (1 - \kappa)\,{\sf K}\right]\;
\frac{t}{{\sf K}} \nonumber \\
 & \equiv & Z(t|\kappa) \;+\; \frac{1}{4}\;J_{\ell}\,\zeta_{\ell},
\label{eq:defs}
\end{eqnarray}
where we inserted the definition \eqref{eq:J_lib} for the librating action $J_{\ell}$. When substituted into Eq.~\eqref{eq:S_first_lib}, we obtain the final expression for the generating function
\begin{equation}
S_{\ell}(t,\kappa) \;=\; 4\,Z(t|\kappa).
\label{eq:S_sol_lib}
\end{equation}
Hence, we see that the Jacobi zeta function $Z(t|\kappa)$ generates the canonical transformation \eqref{eq:canonical_trans} for the librating-pendulum case. In the low-energy limit $(\kappa \ll 1)$, Eq.~\eqref{eq:S_sol_lib} yields $S_{\ell} \simeq \kappa\,\sin 2\,\zeta_{\ell}$, where $J_{\ell} \simeq 2\,\kappa$ and $\zeta_{\ell} \simeq t$, which satisfies the conditions 
\eqref{eq:S_zeta_def}-\eqref{eq:S_J_def} with Eq.~\eqref{eq:thetap_low}.

Lastly, the Jacobi zeta function \eqref{eq:Zeta_def} has the following partial derivatives
\begin{eqnarray}
\pd{\rm Z}{t} & = & {\rm dn}^{2} \;-\; \frac{\sf E}{\sf K}, \label{eq:partial_Z_t} \\
\pd{\rm Z}{\kappa} \;+\; \frac{1}{\Omega_{\ell}}\;\pd{\rm Z}{t} & = & \frac{{\rm cn}}{2\,(1 - \kappa)} \left( {\rm dn}\,{\rm sn} \;-\frac{}{} {\rm Z}\;{\rm cn} \right), \label{eq:partial_Z_kappa}
\end{eqnarray}
where $\Omega_{\ell}$ is defined in Eq.~\eqref{eq:Omega_lib_def}. These partial derivatives can be used to show that the generating function \eqref{eq:S_sol_lib} satisfies the partial derivatives \eqref{eq:S_zeta_def}-\eqref{eq:S_J_def}.

\subsection{Rotation Case}

Next, we consider the rotating-pendulum case, represented by Eqs.~\eqref{eq:theta_rot_zetaJ}-\eqref{eq:p_rot_zetaJ},
\eqref{eq:vartheta_t_rot}-\eqref{eq:p_t_rot}, and \eqref{eq:vartheta_J_rot}-\eqref{eq:p_J_rot}. Equation \eqref{eq:S_zeta_def} for the rotation case can be explicitly written as
\begin{equation}
\pd{S_{r}}{\zeta} \;=\; \frac{\sqrt{\kappa}}{\omega_{r}}\;\pd{S_{r}}{u} \;=\; \frac{4\kappa}{\omega_{r}}\;\ov{\rm dn}^{2} \;-\; J_{r},
\label{eq:S_zeta_rot}
\end{equation}
where Eqs.~\eqref{eq:p_rot_zetaJ} and \eqref{eq:vartheta_t_rot} were used on the right side. Equation \eqref{eq:S_zeta_rot} can be integrated with respect to $u$ to give
\begin{eqnarray}
S_{r}(u,\kappa) & = & 4\,\sqrt{\kappa}\;\int_{0}^{u}\,{\rm dn}^{2}\left(s|\kappa^{-1}\right)\,ds \;-\; J_{r}\,\zeta_{r},
\label{eq:S_first_rot}
\end{eqnarray}
where we assumed that $S_{r} = 0$ at $u = 0$.

Next, we use the definition \eqref{eq:Zeta_def} for the Jacobi zeta function to obtain
\begin{eqnarray}
\int_{0}^{u}\;{\rm dn}^{2}\left(s|\kappa^{-1}\right)\,ds & = & Z\left(u|\kappa^{-1}\right) \;+\; 
\frac{\ov{\sf E}}{\ov{\sf K}}\;u \nonumber \\
 & = & Z\left(u|\kappa^{-1}\right) \;+\; \frac{J_{r}\,\zeta_{r}}{4\,\sqrt{\kappa}},
\label{eq:defs_rot}
\end{eqnarray}
where we used Eqs.~\eqref{eq:J_rot}, \eqref{eq:omega_rot}, and \eqref{eq:zeta_rot}. When this expression is inserted into Eq.~\eqref{eq:S_first_rot}, we obtain the final expression for the generating function for the rotation case 
\begin{equation}
S_{r}(u,\kappa) \;=\; 4\,\sqrt{\kappa}\;Z\left(u|\kappa^{-1}\right).
\label{eq:S_sol_rot}
\end{equation}
Hence, like the libration case [Eq.~\eqref{eq:S_sol_lib}], the Jacobi zeta function plays a fundamental role in generating the canonical transformation \eqref{eq:canonical_trans} for the rotation case.

Lastly, the Jacobi zeta function \eqref{eq:Zeta_def_u} has the following partial derivatives
\begin{eqnarray}
\pd{\ov{\rm Z}}{u} & = & \ov{\rm dn}^{2} \;-\; \frac{\ov{\sf E}}{\ov{\sf K}}, \label{eq:partial_Z_u} \\
\pd{\ov{\rm Z}}{\kappa} \;+\; \frac{1}{\Omega_{r}}\;\pd{\ov{\rm Z}}{u} & = & \frac{\ov{\rm cn}}{2\,\kappa\,(\kappa - 1)} 
\left( \ov{\rm Z}\;\ov{\rm cn} \;-\frac{}{} \ov{\rm dn}\,\ov{\rm sn} \right), \label{eq:partial_Z_kappa_u}
\end{eqnarray}
where $\Omega_{r}$ is defined in Eq.~\eqref{eq:Omega_rot_def}. These partial derivatives can be used to show that the generating function \eqref{eq:S_sol_rot} satisfies the partial derivatives \eqref{eq:S_zeta_def}-\eqref{eq:S_J_def}.

\section{Summary}

The problem of the motion of a pendulum represents a fundamental paradigm in mathematical physics. It is well known that its solution is intimately connected with the Jacobi elliptic functions, which represents an important class of mathematical functions that find applications in physics. 

In the present paper, we have shown that yet another Jacobi elliptic function, the Jacobi zeta function ${\rm Z}(t|\kappa)$, appears naturally in the canonical transformation \eqref{eq:canonical_trans} that defines the action-angle coordinates for the pendulum problem. Indeed, it is used to generate the canonical transformation for the libration motion \eqref{eq:S_sol_lib} and the rotation motion \eqref{eq:S_sol_rot} of the pendulum.

Lastly, the Jacobi-elliptic formulation of the canonical transformation to action-angle coordinates for the pendulum problem can now be applied to the formulation of the canonical transformation to bounce-center action-angle coordinates, which describe the guiding-center trapped/passing particle orbits in axisymmetric tokamak geometry \cite{Brizard_gc}.


\begin{thebibliography}{99}
\bibitem{footnote} We use the notation ${\rm sn}(t|\kappa) \equiv {\rm sn}(t,k)$, where $\kappa = k^{2}$.
\end{thebibliography}
\end{document}